\documentstyle[preprint,prd,aps,floats,eqsecnum,epsf]{revtex}

\newcommand{\bra}[1]{\langle{#1}\vert}
\newcommand{\ket}[1]{\vert{#1}\rangle}

\newcommand{\cg}{{C}_{\mbox{g}}}

\newcommand{\cl}{\cal {L}}

\newcommand{\dd}{\partial}
\newcommand{\dmuup}{\dd^{\mu}}
\newcommand{\dmud}{\dd_{\mu}}
\newcommand{\dnud}{\dd_{\nu}}

\newcommand{\dsf}{\displaystyle\frac}

\newcommand{\expon}[1]{{e}^{\disc{#1}}}

\newcommand{\fpi}{f_{\pi}}

\newcommand{\half}{\displaystyle \frac{1}{2}}

\newcommand{\integ}{\int_{0}^{\infty}dx}

\newcommand{\ltwo}{{\cl}_{\mbox{2}}}

\newcommand{\lfora}{{\cl}_{\mbox{4}}}

\newcommand{\lkin}{ {\cl}_{{\disc\mbox{kin}}} }

\newcommand{\ra}{\rightarrow}

\newcommand{\msig}{m_{\sigma}}

\newcommand{\mpi}{{\disc{m}_{\pi}}}

\newcommand{\nwl}{\\[2mm]}

\newcommand{\pal}{\partial}

\newcommand{\pirow}{\pi\rho\omega}
\newcommand{\pirows}{\pi\rho\omega\sigma}

\newcommand{\pr}[1]{{#1}^\prime}

\newcommand{\sigmat}{\tilde{\sigma}}

\newcommand{\Tr}{\mbox{Tr\,}}

\newcommand{\vecr}{\vec{r}}

\newcommand{\xpr}{\pr{x}    }

\newcommand{\be}{\begin{equation}}
\newcommand{\ee}{\end{equation}}
\newcommand{\beq}{\begin{equation}}
\newcommand{\eeq}{\end{equation}}
\newcommand{\ba}{\begin{array}{l}}
\newcommand{\ea}{\end{array}}
\newcommand{\bea}{\begin{eqnarray}}
\newcommand{\eea}{\end{eqnarray}}
\newcommand{\banonum}{\begin{eqnarray*}}
\newcommand{\eanonum}{\end{eqnarray*}}
\newcommand{\banum}{\begin{eqnarray}}
\newcommand{\eanum}{\end{eqnarray}}
\newcommand{\lab}[1]{\label{#1}}
\newcommand{\bfr}{\begin{flushright}}
\newcommand{\bfl}{\begin{flushleft}}
\newcommand{\efl}{\end{flushleft}}
\newcommand{\efr}{\end{flushright}}
\newcommand{\bi}[1]{\bibitem{#1}}
\newcommand{\bb}{\begin{thebibliography}}
\newcommand{\eb}{\end{thebibliography}}
\newcommand{\bc}{\begin{center}}
\newcommand{\ec}{\end{center}}
\newcommand{\ci}[1]{\cite{#1}}
\newcommand{\disc}{\displaystyle}
\newcommand{\ds}{\displaystyle}
\newcommand {\bdc}{\begin{document}}
\newcommand{\edc}{\end{document}}
\newcommand{\re}[1]{(\ref{#1})}
\renewcommand\thesection{\Roman{section}}
\renewcommand\thesection {\arabic{section}.}
\newcommand{\vecnab}{\mbox{\boldmath $\nabla$}}
\newcommand{\veck}{{\bf k}}
\newcommand{\vecV}{{\bf V}}
\newcommand{\vecB}{{\bf B}}
\newcommand{\tochka}{\, .}
\newcommand{{\vergul}}{\, ,}
\newcommand{\pcal}{\mathcal{{P}}}
\newcommand{\vpcal}{\vec{\mathcal{{P}}}}
\renewcommand\thesection{\Roman{section}}
\newcommand{\doublespace}
{\renewcommand{\baselinestretch}{1.6}\large\normalsize}
\tightenlines
\begin{document}
\draft
\title{\Large\bf {
The sigma - nucleon form  factor  in a chiral topological model
with anomalous  dimension.  }\\
}
\author{
 Abdulla Rakhimov \thanks
{E-mail: rakhimov@rakhimov.ccc.uz
 }}
 \medskip
\address{
Institute of Nuclear Physics, Tashkent, Uzbekistan\\}
\maketitle
\begin{abstract}
\medskip
\medskip
\medskip

The chiral soliton model including $\pi, \rho, \omega$
mesons is extended  for the case when
 the scalar - isoscalar $\sigma$ 
 meson has an anomalous dimension - d. 
  The form factor of the sigma - nucleon 
  interaction    calculated and to be found
  not highly sensitive
  to the value of d.

\end{abstract}
\medskip
\medskip
\newpage

\section{Introduction}
\indent

It is well known that, the scalar isoscalar $\sigma$-meson plays
an important role in nucleon - nucleon (NN) interaction. In one boson exchange 
model (OBE) of NN interaction one has to take into account 
the form - factor of $\sigma NN $  vertex which is still
poorly known. In our recent paper \ci{ourpirho} we considered the form - factor
in the framework of $\pirows$  soliton model including a dilaton
field as a sigma meson with the ordinary scale dimension $d=1$.
On the other hand, there are some versions  of chiral effective Lagrangians 
for the finite nuclear matter and finite nuclei 
where   light scalar
 degrees of freedom have  an anomalous scale dimension \ci{fts}.
 
 Note that, anomalous, fractal, scale dimensions are ingredient
 feature of modern theories. In particular it has been
 recently shown that,  fractal scale dimension
 is essential to describe   superconductivity \ci{ourprb}.
  In present brief note we  study the role of anomalous
   scale dimension in $\sigma NN$ interaction.  
 The paper is organized as follows. In Sect. II we outline 
 the  scaling prpoperties of basic fields; in Sect. III we shall extend the 
 $\pi,\rho,\omega$ Lagrangian for the case of anomalous  
 dimension; in Sect. IV
  we  calculate the vertex form factor of $\sigma N$ interaction.
  The results and brief summary  are presented in Sect. V

\section{Scaling behavior of scalar and spinor fields}
\indent

In this section, to make the point clearer and for further
 references we briefly discuss 
the scaling properties of various fields. For illustration 
we consider kinetic terms of fermion and boson fields.
 The kinetic terms of
a fermion field (e.g. quark field in QCD, or nucleon field)
is given by 
\be
\lkin^{\psi}(x)=i\bar\psi (x)\dmuup \gamma_\mu \psi (x)
\ee
 and a boson ($\pi$ or $\sigma$) field
\be
\lkin^\pi(r)=\half \dmud\pi \dmuup\pi .
\ee
 The scale invariance demands that the action
 $S=\int d^4 x \cal L $ should be invariant
under scale transformations $x_{\mu}=\expon{-\lambda}\xpr_\mu  : $
 \be
S=\int d^4 x \lkin(x)=\int d^4 \xpr\lkin^\prime(\xpr)
\lab{sef}
\ee
\lab{wain5}
Let the fermion and  boson fields scale as
 $\psi(x)=\exp{(n\lambda)}       \psi^\prime(\xpr) $  and
 $\pi(x)=\exp{(m\lambda)}       \pi^{\prime}(\xpr), $ respectively.
For fermions we get
\be
\ba
 S=i\ds\int d^4 x \bar\psi (x)\dmuup \gamma_\mu \psi (x)=
   i\ds\int d^4 \xpr       \bar\psi^{\prime}
 (\xpr)\dmuup \gamma_\mu\psi^\prime (\xpr)
\expon{\lambda}\expon{2n\lambda}\expon{-4\lambda}=
\nwl
=\ds\int d^4 \xpr \lkin^\prime(\xpr)
\ea
\ee
and hence $-4\lambda+\lambda+2n\lambda=\lambda(2n-3)=0 $.
 Since $\lambda$ is arbitrary then $n=3/2$.
 Similarly, for boson fields one can show that $m=1$.
 Therefore we may  conclude
that
\be
x_{\mu}\ra\expon{-\lambda}x_\mu, \quad
\psi(x)\ra\expon{3\lambda/2} \psi(x) ,\quad
 \pi(x)\ra\expon{\lambda} \pi(x)\vergul
\ee
i.e. the scale dimension $d$ equals to $d_f=3/2 $ and $d_b =1$ for
fermions and bosons respectively.
Thus, the term "anomalous
   scale dimension" e.g. for a boson field means
   $d_b \neq 1$. 

In Skyrme like models the pion field is involved 
through a chiral nonlinear field $U=\exp{(i\vec\tau \cdot \vec\pi/\fpi)}$ which
has trivial  scale dimension $d_U=0$. 
For this reason the four derivative term
of the Skyrme Lagrangian  given by
\be
\lfora=\dsf{1}{32 e^2}\Tr[(\dmud U)U^{+},(\dnud
U)U^{+}]^2
\ee
 is scale invariant by itself:
\be
\int d^4 x\lfora(x)\ra\int d^4\xpr\expon{-4\lambda}      \lfora^\prime
(\xpr)
\expon{4\lambda}= \int d^4 \xpr      \lfora^\prime  (\xpr)\vergul
\ee
whereas the kinetic term
\be
\ltwo(x)=\dsf{\fpi^2}{4}\Tr [\dmud U(x)\dmuup U^{+}(x)]
\ee
is not:
\be
\ba
\ds\int d^4 x\ltwo(x)\ra \dsf{\fpi^2}{4}
\ds\int d^4\xpr\expon{-4\lambda}\expon{2\lambda}
\Tr [\dmud        U^\prime (\xpr)\dmuup U^{'+}(\xpr)]=
\nwl
=\expon{-2\lambda}\ds\int d^4 \xpr      \ltwo^\prime (\xpr)
\ea
\ee
How can the invariance  be restored?
According to Shehter \ci{shehter}
an additional boson field $\chi(x) $ should be included into
the Skyrme Lagrangian. We shall use this prescription in the next Section.

\section{The $\pirows$  model with anomalous dimension}
\indent

 We start with extended  Skyrme - like Lagrangian including
 vector mesons explicitely \ci{meisnpa}
\be
\ba
{\cl}=
-\dsf{f_{\pi}^{2}}
{4}\Tr L_{\mu}L^{\mu}
-\dsf{f_{\pi}^{2}}{2}
\Tr[l_{\mu}+ r_{\mu}+ig\vec\tau\vec\rho_{\mu}
+ig\omega_{\mu}]^2
+\dsf{3}{2}g\omega_{\mu}B^{\mu}-
\dsf{1}{4}(\omega_{\mu\nu}\omega^{\mu\nu}+
\vec\rho_{\mu\nu}\vec\rho^{\, \, \mu\nu})
\nwl
\\
\quad\quad+\dsf{f_{\pi}^2m_{\pi}^{2} }{2}\Tr(U-1)\vergul
\label{anom0}
\ea
\ee
where 
 left/right--handed currents are given by
$L_{\mu}=U^+\pal_{\mu}U$, $l_{\mu}=\xi^+\pal_{\mu}\xi$,
  $ r_{\mu}=\xi\pal_{\mu}\xi^+$,$  \quad\xi=\sqrt{U}$,
and the pertinent vector meson ($\vec{\rho}, \omega$) field strength tensors
are
$\vec\rho_{\mu\nu}=\pal_{\mu}\vec\rho_{\nu}-\pal_{\nu}\vec\rho_{\mu}+
g[\vec\rho_{\mu}\times\vec\rho_{\nu}]$ and
$\omega_{\mu\nu}=\pal_{\mu}\omega_{\nu}-\pal_{\nu}\omega_{\mu}$.
Furthermore, the topological baryon number current is given by
$B^{\mu}=\varepsilon^{\mu\alpha\beta\gamma}
\Tr L_{\alpha}L_{\beta}L_{\gamma}/(24\pi^2)$. Note that, 
for the case of infinite $\rho$ and $\omega$ meson masses the Lagrangian
in Eq. \re{anom0} becomes equal to the usual Skyrme -Witten Lagrangian
with two, four and six derivative terms and, hence, is not  
 scale invariant. As it has been mentioned above this shortcoming can be 
restored by inclusion of a dilaton field - sigma meson. 
Thus, including a  $\sigma$--meson by means of the scale invariance
and trace anomaly of QCD into the Lagrangian
one obtains following  chiral Lagrangian of the coupled $\pirows$
system
\be
\ba
{\cl}=\dsf{ S_{0}^{2}e^{-2\sigma/d} }{2 } \pal_{\mu}\sigma
\pal^{\mu}\sigma
-\dsf{f_{\pi}^{2}e^{-2\sigma/d }}
{4}\Tr L_{\mu}L^{\mu}
-\dsf{f_{\pi}^{2}e^{-2\sigma/d} }{2}
\Tr[l_{\mu}+ r_{\mu}+ig\vec\tau\vec\rho_{\mu}+
\nwl
\\
+ig\omega_{\mu}]^2
+\dsf{3}{2}g\omega_{\mu}B^{\mu}-
\dsf{1}{4}(\omega_{\mu\nu}\omega^{\mu\nu}+
\vec\rho_{\mu\nu}\vec\rho^{\, \, \mu\nu})
+\dsf{f_{\pi}^2m_{\pi}^{2}e^{-3\sigma/d} }{2}\Tr(U-1)-
\nwl
\\
-\dsf{d^{2}S_{0}^2\msig^{2}}{16}[1-e^{-4\sigma/d}(\dsf{4\sigma}{
d}+1)], \\
\label{anom1}
\ea
\ee
Here the scale dimension of $\sigma$ field ($d\geq1$) is included explicitly.
In  Eq.\re{anom1} $S_0$ is the vacuum expectation value of scalar field
in free space matter,
$\fpi$ is the pion decay constant $(\fpi=93~{\rm MeV})  $
 and $g=g_{\rho\pi\pi}$
 is determined through the KSFR  relation $g=m/\sqrt{2}\fpi $.
 The model assumes the masses
of $\rho$ and $\omega$ mesons to be equal, $m_\rho=m_\omega=m$.
The mass of the $\sigma$ is related
to the gluon condensate in the usual way
 $\msig=2\sqrt{\cg}/(dS_0) ~\cite{fts}$.

Nucleons arise as soliton solutions  from the Lagrangian Eq.\re{anom1}
in the sector with baryon number $B=1$ .
To construct them one goes through a two step procedure. First,
one finds the classical soliton which has
neither good spin nor good isospin.
Then an adiabatic rotation of the soliton is performed
and it is quantized collectively.
The classical soliton follows from Eq.\re{anom1}
by virtue of a spherical symmetrical ans\"atze for the meson fields:
\be
\ba
U(\vec r)=\exp{(i\vec\tau\hat{r}\theta(r))},  \, \, \,
\quad \rho_i^a=\varepsilon_{iak}\hat r_k\dsf{G(r)}{gr},
\quad \omega_{\mu}(\vec r)=\omega(r)\delta_{\mu 0},  \, \, \,
\sigma(\vec r)=\sigma(r)\,\,.
\lab{anom2}
\ea
\ee
In what follows we call $\theta(r) $,  $G(r) $,
 $\omega(r) $,  and $\sigma(r)  $
the pion--, $\rho $--,  $\omega $--,  and $\sigma$--meson
profile functions, respectively.
The pertinent boundary conditions to ensure baryon number one
and finite energy are,
$\theta(0)=\pi,  G(0)=-2,  \omega^{\prime}(0)=\sigma^{\prime}(0)=0,
\theta(\infty)= G(\infty)=\omega(\infty)=\sigma(\infty)=0$.
To project out baryonic states of good spin and isospin,  we
perform a time--independent SU(2) rotation
\be
\ba
U(\vecr, t)=A(t)U(\vecr)A^{+}(t), \, \,
 \xi(\vecr, t)=A(t)\xi(\vecr)A^{+}(t)\\
 \sigma(\vecr, t)=\sigma(r), \quad
\omega(\vecr, t)=\dsf{\phi(r)}{r}[\vec K \hat r]\\
\vec\tau\cdot
\vec\rho_{0}(\vecr, t)=\dsf{2}{g}A(t)\vec\tau\cdot(\vec K\xi_1(r)+
\hat r\vec K\cdot\hat r \xi_2(r))A^+(t), \\
\vec\tau\cdot\vec\rho_{i}(\vecr, t)=A(t)\vec \tau
\cdot\vec\rho_{i}(\vecr)A^{+}(t)
\lab{anom3}
\ea
\ee
with $2\vec{K}$ the angular frequency of the spinning mode of soliton,
$i\vec\tau\cdot\vec K=A^{+}\dot{A}$.
This leads to the time--dependent Lagrange function
\be
{\cal L}(t)=\int d\vecr {\cal L}=-M_{H}(\theta, G, \omega, \sigma)
+\Lambda(\theta, G, \omega, \sigma, \phi, \xi_1, \xi_2)\Tr(\dot{A}\dot{A}^+)~.
\ee
Minimizing the classical mass $M_{H}(\theta, G, \omega, \sigma)$
 leads to the coupled differential equations
for $\theta, G, \omega$ and $\sigma $
 subject to the aforementioned boundary conditions.
In the spirit of the large $N_c $--expansion,  one then extremizes the
moment of inertia $\Lambda(\theta, G, \omega, \sigma, \phi, \xi_1, \xi_2) $
 which gives the coupled differential
equations for $\xi_1$ , $ \xi_2  $ and $\phi $
 in the presence of the background profiles
$\theta,  G,  \omega $ and $\sigma  $.
  The pertinent boundary conditions are
$\phi(0)=\phi(\infty)=0, \, \,$
$\xi_1^{\prime}(0)=\xi_1(\infty)=0,\,\,$
$\xi_2^{\prime}(0)=\xi_2(\infty)=0,\,\,  $
$2\xi_1(0)+\xi_2(0)=2. $
The masses of nucleon $M_N $ and the mass of $\Delta$ ,
 $M_{\Delta} $,
are then given by $M_N=M_H+3/8\Lambda $ and $M_\Delta=M_H+3/15\Lambda$.
\section{Vertex form factor of $\sigma N$ interaction}
\indent

The meson - nucleon form factors can be derived by a well known procedure,
proposed long years ago by Cohen~\ci{cohen,meisff}.
 Although they were derived in a microscopical
and consequent way, these form factors could not be directly used
in standard OBE schemes of nucleon - nucleon interaction.
 The reason is that the OBE schemes
 ~\ci{machrep} in momentum space  use form factors
defined for the fields propagating on a flat metric, whereas
the definition of form factors in Cohen's procedure
involve a nontrivial
metric. Hence, before using  them in OBE scheme
one should modify the procedure by redefining meson fields.
Note that, the modification for pion nucleon form factors
in $\pirow$ model is clearly outlined in refs. ~\ci{ourpirho,holzmach1}.

Redefinition of a meson field i.e. introduction of a flat metric 
is based on a canonical form for the kinetic part of the Lagrangian,
which determines the dynamics of the field fluctuation.
The kinetic term
of the scalar meson in Eq. \re{anom1}
\be
{\cal L^{\mbox{kin}}_{\sigma}}= S_{0}^{2}e^{-2\sigma/d}  \pal_{\mu}\sigma
\pal^{\mu}\sigma/2
\ee
 can be easily rewritten in a  usual way:
\be
{\cal L^{ \mbox{kin}  }_{\sigma}}= \pal_{\mu}\sigmat\pal^{\mu}\sigmat/2
\ee
by the following redefinition of the basic sigma field:
\be
\sigmat(r)=S_0d[1-e^{-\sigma(r)/d}].
\lab{sigmat}
\ee
 Now the new field
$\sigmat$ may be identified with a real sigma field. It
can be also easily shown that the above redefinition does not 
modify the mass: $m_{\sigmat}=m_{\sigma}$.

Now we apply Cohen's procedure to derive $\sigma NN$ form factor.
Klein - Gordon equation for the  scalar-isoscalar meson interacting with nucleon 
with the Lagrangian 
${\cal L}_{\sigma NN}=g_{\sigma NN}\bar\psi\psi {\sigmat}$
is given by:
\be
 (-\vec \nabla^2 +m_{\sigma}^{2}\sigmat)=j
\lab{kg}
\ee
with the source  $j(x)=g_{\sigma NN}\bar\psi(x)\psi(x)$.
The first step of the procedure assumes taking matrix elements of
both sides of Eq. \re{kg} between nucleon states $N(\vpcal')$ and $N(\vpcal)$.
Evaluating the matrix element of the source
in  Breit frame, $(\vpcal'=-\vpcal$,
$\vec q=\vpcal -\vpcal')$ one can find
\be
\ba
\bra{N(\vpcal')}\vec j(r)\ket{N(\vpcal)}=
\dsf{
\exp(i\vec q\,\vec r)
}
{(2\pi)^{3}}
\bra{N(\vpcal')}\vec j(0)\ket{N(\vpcal)}
=G_{\sigma NN}(t)\bar{u}(\vpcal')u(\vpcal)\nwl
=\dsf{E_q}{2M}[1+\dsf{\vec{q}^{\ 2}}{4(E_q+M)^2}       ]
G_{\sigma NN}(t)\equiv G_{\sigma NN}(t)T(\vec{q}^{\ 2}),
\lab{melj}
\ea
\ee 
where  $t=-\vec{q}^{\ 2}$, and $u(p)$ is Dirac spinor 
\begin{eqnarray}
u(\vec p)=\left(\dsf{E_p+M}{2M}\right)^{1/2}
\left(\begin{array}{c}
1\\
(\vec \sigma\vec p)/(E_p+M)
\lab{up}
\end{array}\right)\chi_S,
\end{eqnarray}
with the normalization $\bar u(p)u(p)=1$.
The similar matrix element for the $\sigma $ field with collective coordinates
$r$ is given by:
\be
\ba
\bra{N(\vpcal')}\sigmat(\vecr-\vec{R})\ket{N(\vpcal)}=
\dsf{1}{(2\pi)^{3}}\displaystyle\int d\vec R \exp(i\vec q\,\vec R)
\bar{u}(\vpcal')\sigmat(\vecr-\vec{R})u(\vpcal)\nwl
=-\dsf{\exp(i\vec q\,\vec r)T(\vec{q}^{\ 2})}{(2\pi)^{3}}
\displaystyle\int d\vec x \exp(-i\vec q\,\vec x)
\sigmat(\vec x),
\lab{melsig}
\ea
\ee
where $\vec R=\vecr-\vec x$. Now using Eqs. \re{melj} and \re{melsig}
in Eq. \re{kg} one obtains
\be
 \displaystyle\int d\vec x \exp(-i\vec q\,\vec x)
 (\vec{q}^{\ 2}+m_{\sigma}^{2})\sigmat(\vec x)=-G_{\sigma NN}(t).
\lab{qq}
\ee
For the spherical symmetric ansatz with $\sigmat(\vec x)=\sigmat( |x|)$
the left hand side of Eq. \re{qq} can be easily evaluated:
\be
\ba
\displaystyle\int d\vec x \exp(-i\vec q\,\vec x)
 (\vec{q}^{\ 2}+m_{\sigma}^{2})\sigmat(x)=
 \displaystyle\int d\hat{x}  dx \; x^2\exp(-i\vec q\,\vec x)[-\vec \nabla^2 +m_{\sigma}^{2}]
 \sigmat(x)\nwl
 \\
 =4\pi\displaystyle\integ  
 \{ -j_0(qx)[2x\sigmat'(x)+x^2\sigmat''(x)]+j_0(qx)x^2m_{\sigma}^{2}\sigmat(x) \}
 \vergul
\ea
\ee
where $j_0(x)$ - spherical Bessel function.
 
Finally, inserting this into the Eq. \re{qq} we get following
expression for the $\sigma NN$ form factor
\be
G_{\sigma NN}(t)=4\pi\displaystyle\integ j_0(qx)  
 \{ x^2\sigmat''(x)+2x\sigmat'(x)-x^2m_{\sigma}^{2}\sigmat(x) \}
 \lab{ff}
 \ee
 Note that, $\sigmat (x)$ is defined in Eq. \re{sigmat}, where 
 $\sigma (x)$ is the solution of the coupled nonlinear equations
 of the motion, obtained by minimizing $M_H$. 

   \section{Results and summary.}
\indent

 Before doing
actual calculations the parameters of the model
should be clarified.
As can be seen from Eq.\re{anom1}, the Lagrangian
has no free parameters in the $\pirow$ sector. So,  in actual calculations
the parameters $\mpi,  m,  \fpi$ are fixed at their emperical values,
$\mpi=138$~MeV, $m=m_{\rho}=m_{\omega}=770$~MeV,  $\fpi=93$~MeV,
$ g=m/\sqrt{2}\fpi=5.85$.  In the $\sigma$--meson sector there are
in general three  free parameters:  $\msig$, $S_0$ and
the  scale dimension $d$. 
The values for $S_0$ - were  found to be $S_0=90.6\div 95.6$  MeV ~\ci{fts} .
 So we put  $S_0=\fpi=93$  MeV. As to the mass of sigma meson 
 $\msig$ we set $\msig=550 MeV $ to get the best description of
 static nucleon properties. To study the role of the scale dimension
  we shall consider two cases: $d=1$ and $d=2.4$ corresponding to
 the normal and anomalous cases respectively.

The $\sigma NN$ form factor $G_{\sigma NN}(t)$ given in Eq. \re{ff}
is presented in
 Fig.1.  The solid
and dashed curves are  for $d=2.4$ and $d=1$ cases respectively.
 It is seen that
both curves
coincide qualitatively, so that, being used in 
OBE model they would lead to the same NN potential.
Thus, the sensitivity of $G_{\sigma NN}(t)$
to the value of $d$ need  not to be large. 

We  point
out that    the sigma meson--nucleon form factor found in the present
model could be useful in a wider context of calculations
of nucleon--nucleon observables (phase shifts, deuteron properties
etc) and may give  more information on meson--nucleon
and nucleon--nucleon dynamics.

To summarize,  we have developed a topological chiral soliton model
with an explicit light scalar--isoscalar meson field with anomalous dimension,
 which
plays a central role in nuclear physics, based on the
chiral symmetry and broken scale invariance of QCD. 
We have shown that, the anomalous
value of scale dimension would not lead to substantial modification of
the $\sigma NN$ vertex form - factor.  
\section*{Acknowledgments}
This work is supported by Uzbek Science Foundation $\Phi-2.1.25$.

\begin{figure}
 \epsfclipon
 \epsfxsize=18.cm
 \epsfysize=19.4cm
\epsffile{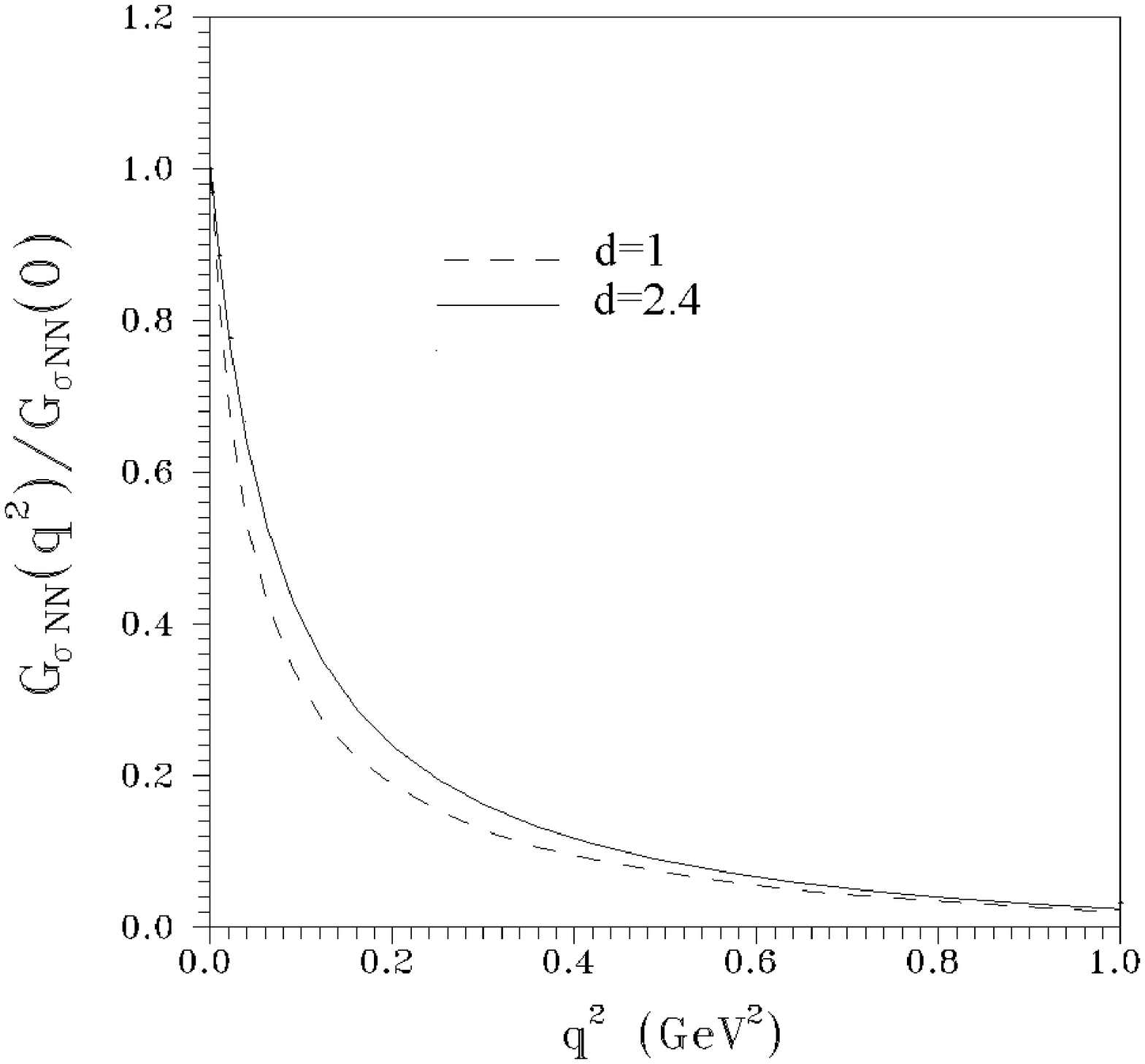}
\caption{The normalized  $\sigma NN$ form factor, $G_{\sigma NN}(t)$.
The solid and dashed lines are for 
$d=2.4$ and $d=1$ cases respectively.
}
\end{figure}
\edc